\documentclass[fleqn,usenatbib]{mnras}

\usepackage{newtxtext,newtxmath}

\usepackage[T1]{fontenc}

\DeclareRobustCommand{\VAN}[3]{#2}
\let\VANthebibliography\thebibliography
\def\thebibliography{\DeclareRobustCommand{\VAN}[3]{##3}\VANthebibliography}

\usepackage{graphicx}	
\usepackage{amsmath}	
\usepackage{amssymb}	

\title[X-ray spectra from compact objects]{Observing imprints of black hole event horizon on X-ray spectra}

\author[S. Banerjee et al.]{Srimanta Banerjee$^{1}$\thanks{Contact e-mail: \href{mailto:srimanta.banerjee4@gmail.com}{srimanta.banerjee4@gmail.com}}, Marat Gilfanov$^{2, 3}$, Sudip Bhattacharyya$^{1}$ and Rashid Sunyaev$^{2,3}$\\
$^{1}$Department of Astronomy and Astrophysics, Tata Institute of Fundamental Research, 1 Homi Bhabha Road, Mumbai 400005, India\\
$^{2}$Max Planck Institute for Astrophysics, Karl-Schwarzschild-Strasse. 1, Garching b. Munchen D-85741, Germany\\
$^{3}$Space Research Institute of Russian Academy of Sciences, Profsoyuznaya 84/32, 117997 Moscow, Russia\\
}

\date{Accepted 2020 September 7. Received 2020 September 7; in original form 2020 April 24}

\pubyear{2020}

\begin{document}
\label{firstpage}
\pagerange{\pageref{firstpage}--\pageref{lastpage}}
\maketitle

\begin{abstract}
A fundamental difference between a neutron star (NS) and a black hole (BH) is the absence of a physical surface in the latter.  For this reason,  any remaining  kinetic energy of the matter accreting onto a BH is advected inside its event horizon. In the case of an NS, on the contrary,  accreting material is decelerated on the NS surface, and its kinetic energy is eventually radiated away.  Copious soft photons produced by the NS surface will affect the properties of the Comptonised component dominating spectra of X-ray binaries in the hard state. Thus, parameters of the Comptonised spectra --  the electron temperature $kT_{\rm e}$ and the Compton $y$-parameter, could serve as an important tool for distinguishing BHs from NSs. In this paper, we systematically analyse heretofore the largest sample of spectra from the BH and NS X-ray binaries in the hard state for this purpose, using archival \textit{RXTE}/PCA and \textit{RXTE}/HEXTE observations.  We find that the BHs and NSs occupy distinctly different regions in the $y-kT_{\rm e}$ plane with NSs being characterised by systematically lower values of $y$-parameter and electron temperature. Due to the shape of the boundary between BHs and NSs on the $y-kT_{\rm e}$ plane, their one-dimensional $y$ and $kT_{\rm e}$ distributions have some overlap.  A cleaner one parameter diagnostic of the nature of the compact object in X-ray binaries is provided by the Compton amplification factor $A$, with the boundary between BHs and NSs lying at $A\approx 3.5-4$.
This is by far the most significant detection of the imprint of the event horizon on the X-ray spectra for stable stellar-mass BHs.

\end{abstract}




\section{Introduction}\label{intro}
It has been known for decades that the broad-band energy spectra of black hole X-ray binaries (BHXBs), which harbour accreting stellar-mass black hole candidates (hereafter, we will drop `candidate'), can be broadly divided into optically thick and optically thin spectral components \citep{Remillard, Done2007, gilfanov2010}. The optically thick black body-like component is well described as originating from a geometrically thin, optically thick accretion disc \citep{Sunyaev73}  and it typically has a colour temperature in the range of $kT_{\rm bb}\sim 0.1 - 1$ keV \citep{Remillard, gilfanov2010} ($k$ is the Boltzmann's constant). The optically thin component exhibits roughly a power law shape from a few keV to $\sim$ hundred keV followed by  a roughly  exponential cut-off \citep{Syunyaev91, Remillard,gilfanov2010}, and is produced due to the Compton up-scattering of seed photons by a hot optically thin electron cloud of temperature $kT_{\rm e}$ and optical depth $\tau$, located somewhere near the black hole \citep{Sunyaev79,Sunyaev80}. Depending upon the relative dominance of these two components, the X-ray emission states of BHXBs can be broadly divided into two classes: soft (or high) state, when the thermal disc component is dominant and hard (or low) state, when the hard Comptonised component is dominant \citep{Remillard,gilfanov2010}. The X-ray spectrum of BHXBs also exhibits reflection features, which are generated as a fraction of the Comtonised photons get scattered or re-emitted into the line of sight by the accretion disc. The most prominent features of this spectral component are an iron K$\alpha$ emission line near $6.4$ keV, formed due to fluorescence, and a Compton reflection hump peaked around $30$ keV and produced due to the inelastic scattering of the Comptonised photons by free electrons in the disc \citep{Basko1973, Ross2005}.

A subclass of  accreting neutron stars, with low magnetic field ($B \sim 10^{8-9}$~G), also exhibits the above mentioned spectral properties \citep{barret2001, Done,gilfanov2010}. These are the atoll sources (including millisecond pulsars), which show a transition between the hard `island' state and the soft `banana' state \citep{klis89}, equivalent to spectral state transitions in black holes. The neutron star radii are of the order of $\sim3R_S$ ($R_S=2GM/c^2$ is the Schwarzschild radius; $M$ is the mass of the compact object, and $c$ is the speed of light in vacuum), which is comparable to the radius of the last marginally stable orbit around a black hole \citep{Done2007}. Outside a few Schwarzschild radii, black holes and neutron stars  have similar shapes of the gravitational potential well. As a consequence, one may expect that at the same values of $\dot{M}/\dot{M}_{\rm Edd}$ ($\dot{M}$ is the mass accretion rate  and $\dot{M}_{\rm Edd}$ is the critical  Eddington mass accretion rate), the physical structure of accretion flow around a black hole should be similar to that for a neutron star. 

However, there exists a fundamental difference between a black hole and a neutron star: the lack of a physical surface in the case of the former, while the latter has a surface. Such a difference gives rise to  prominent observable effects \citep{Syunyaev91} like thermonuclear bursts due to the unstable nuclear burning of the accreted material on the surface of neutron stars \citep{narayan2002}, or  differences in the spectral evolution of neutron star and black hole transients \citep{Done}, or pulsation from the residual magnetic field \citep{Done2007}. On the other hand, a boundary or spreading layer \citep{Sunyaev88, Inogamov1999}  is formed between the inner part of the accretion flow moving with the Keplerian velocity (or some fraction of it) and the surface of the more slowly rotating neutron star.  In this layer, the accreting material is decelerated down to the velocity of the neutron star surface, and its kinetic energy is deposited \citep{gilfanov2014}. In Newtonian geometry, half of the gravitational potential energy of the accreting matter is released in the boundary layer (the other half being radiated away from the disc), while this fraction is even greater in General relativity \citep{Sibgatullin2000}. Thus, the X-ray emission from a neutron star has two constituents of comparable luminosity originating in the accretion disc and boundary layer.
It was shown using the Fourier resolved spectroscopy that at $\dot{M}/\dot{M}_{\rm Edd}>0.1$ the boundary layer emission spectrum is same  in different sources, weakly depends on the mass accretion rate and can be approximately described with a Wien spectrum with a colour temperature of about  $2.4$ keV \citep{Gilfanov}. Note that additional  turbulence expected in the boundary layer, and the resulting variability, make the Fourier resolved spectroscopy a workable tool for neutron stars. For a black hole, on the contrary, the kinetic energy of the accreting material is advected inside the event horizon; therefore, such a luminous soft emission component is absent in black hole X-ray binaries.

In the hard spectral state of black holes and neutron stars, the main spectral formation mechanism is unsaturated Comptonisation \citep{Sunyaev79,Sunyaev80} of soft seed photons in the hot corona located in the vicinity and, presumably, around the compact object.
Additional soft X-ray photons emitted by the surface of  neutron stars can interact  with the hot electrons in the corona  through Comptonisation, significantly changing its energy budget as compared with the black hole case, and leave an imprint on the X-ray spectra of neutron stars in the hard state  \citep{Sunyaev89}. The shape of the Comptonised spectrum is determined  by three quantities: the temperature ($kT_{\rm e}$) of the hot electron cloud or corona, the Compton $y$-parameter of the electron cloud and the temperature of seed soft photons, while the energy balance in the corona is characterized by the Compton amplification factor $A$ which can be derived from the above  quantities \citep{Burke1}. The Compton $y$-parameter (defined as $4kT_e/m_e c^2\cdot Max(\tau,\tau^2)$ with $m_e$ as the rest mass of an electron) describes the average change in energy a collection of photons will suffer as it traverses the Comptonising region. On the other hand, the amplification factor $A$ is determined by the ratio of the energy deposited in the hot electrons to the luminosity of the soft seed photons. Thus, it should be possible to decode the signature of a physical surface or lack of it in the X-ray spectra by studying the distributions of $kT_{\rm e}$, $y$, and $A$ parameter values in the hard state for black hole and neutron star populations. 

Motivated by these arguments, \citet{Burke1} analysed 59 spectra of 7 BH and 5 NS binaries. They found in their sample a  dichotomy in the distribution of $y$-parameter and Compton amplification factor $A$ between black holes and neutron stars, however the electron temperature distributions, although different, spanned an almost similar range of values. In the follow-up paper \citep{Burke2}, analysing a significantly larger number of NS spectra, they found that marginalised one-dimensional distributions of $y$ and $A$ of NS systems may overlap with those for black holes, depending on the NS spin and critical Eddington ratio. In this paper, we investigate this further, aiming to formulate a comprehensive observational picture of the effect of the NS surface on  spectral formation in X-ray binaries in the hard spectral state. To this end,  we  significantly increased the number of sources and spectra. We doubled the sample of objects to include 11 accreting stellar-mass black holes and 13 atoll neutron stars in the hard state and dramatically increased the number of  observations included in our analysis, to nearly 
$5000$. Overall, our strategy is similar to that used in \citet{Burke1}, albeit with some important  modifications detailed below.

The paper is organized as follows. In section \ref{data}, we briefly describe the data reduction schemes, the criteria we set for selecting the \textit{RXTE} observations, and the spectral models we considered for fitting the data. We report our findings and discuss our results in  section \ref{result}. We summarise our results and conclude in  section \ref{summary}.

\section{Data Reduction and Analysis}\label{data}
We collect data ($\sim5000$ observations) for 11 accreting black holes and 13 accreting neutron stars (atoll sources and millisecond pulsars) from the archival \textit{RXTE} database. In the case of neutron stars, we do not consider Z sources \citep{klis89} in this study as we require sources to exhibit classic hard state (`island' state). We use data from both the Proportional Counter Array (PCA) and High-Energy X-ray Timing Experiment (HEXTE) onboard \textit{RXTE}, and those data are reduced according to the procedure mentioned in the \textit{RXTE} cookbook
\footnote{\label{foot:itas1}\url{https://heasarc.gsfc.nasa.gov/docs/xte/recipes/cook_book.html}}. Our data analysis procedure follows, with some modifications,  the approaches proposed in \cite{Burke1}.

\begin{table*}
{
\centering
 \caption{Details of the sources used in this study. The list of sources (the upper ones are black holes and the lower ones are neutron stars) used in this work, along with the no. of observations analysed for each of them $(N_{\rm obs})$, their absorption column densities $(N_{H})$, distances $(D)$ and masses $(M)$ are mentioned in the table. The black hole SAX 1711.6-3808 does not have a well determined mass or distance. For this source, we used ad hoc values of  $10M_{\odot}$ and $5\ \rm kpc$ respectively.  We assume all the neutron stars have the mass of  $1.4M_{\odot}$. References mentioned in the table are: (1) \citep{Steiner}, (2) \citep{Dunn}, (3) \citep{Gladstone}, (4) \citep{Burke1}, (5) \citep{Burke2}, (6) \citep{Munoz}, and (7) \citep{Nathalie}.}.
\medskip
 \begin{tabular}{lllllll}
  \hline
  No. & Source & $N_{\rm obs}$ &\hspace{0.28cm} $N_{H}$ & Distance & Mass & References\\
     &  &  & $(10^{22}\ \rm cm^{-2})$ &\ \rm Kpc & $M_{\odot}$\\
  \hline
  1 & Cyg X-1 & 169 &\hspace{0.28cm} 0.7 & 1.86 & 14.8 &\hspace{0.28cm} (1), (2)\\
  2 & XTE J1752-223 & 52 &\hspace{0.28cm} 0.6 & 3.5 & 9.6 &\hspace{0.28cm} (1), (2)\\
  3 & GX 339-4 & 68 &\hspace{0.28cm} 0.3 & 8.0 & 5.8 &\hspace{0.28cm} (1), (2)\\
  4 & XTE J1550-564 & 24 &\hspace{0.28cm} 0.8 & 4.4 & 10.39 &\hspace{0.28cm} (1), (2)\\ 
  5 & SWIFT J1753.5-0127 & 107 &\hspace{0.28cm} 0.15 & 5.0 & 10.0 &\hspace{0.28cm} (1), (2)\\
  6 & GS 1354-64 & 4 &\hspace{0.28cm} 2.0 & 26.0 & 7.47 &\hspace{0.28cm} (1), (2)\\
  7 & XTE J1650-500 & 2 &\hspace{0.28cm} 0.5 & 2.6  & 4.72 &\hspace{0.28cm} (1), (2)\\
  8 & GRO J1655-40 & 6 &\hspace{0.28cm} 0.7 & 3.2 & 6.6 &\hspace{0.28cm} (1), (2)\\
  9 & SAX 1711.6-3808 & 1 &\hspace{0.28cm} 2.8 & 5.0 & 10.0 &\hspace{0.28cm} (2)\\
 10 & H 1743-322 & 1 &\hspace{0.28cm} 2.2 & 10.4 & 13.3 &\hspace{0.28cm} (1), (2)\\
 11 & XTE J1748-288 & 6 &\hspace{0.28cm} 7.5 & 10.0 & 10.0 &\hspace{0.28cm} (1), (2)\\
  \hline
  1 & 4U 1636-536 & 44 &\hspace{0.28cm} 0.37 & 6.0 & &\hspace{0.28cm} (3), (4)\\
  2 & Aql X-1 & 54 &\hspace{0.28cm} 0.28 & 5.2 & &\hspace{0.28cm} (4)\\
  3 & 4U 1728-34 & 51 &\hspace{0.28cm} 1.24 & 4.6 & &\hspace{0.28cm} (4)\\
  4 & GS 1826-238 & 57 &\hspace{0.28cm} 0.17 & 6.0 & &\hspace{0.28cm} (3), (5)\\
  5 & 4U 1608-52 & 90 &\hspace{0.28cm} 1.6 & 3.3 & &\hspace{0.28cm} (4), (6)\\
  6 & 4U 1724-30 & 45 &\hspace{0.28cm} 1.0 & 6.6 & &\hspace{0.28cm} (3)\\
  7 & HETE J1900.1-2455 & 23 &\hspace{0.28cm} 0.2 & 4.7 & &\hspace{0.28cm} (7)\\
  8 & 4U 0614+09 & 22 &\hspace{0.28cm} 0.45& 3.0 & &\hspace{0.28cm} (3), (5)\\
  9 & 4U 1705-44 & 4 &\hspace{0.28cm} 0.67& 8.4 & &\hspace{0.28cm} (5)\\
  10 & 4U 1820-303 & 1 &\hspace{0.28cm} 0.28 & 5.8 & &\hspace{0.28cm} (3)\\
  11 & Ara X-1 & 4 &\hspace{0.28cm} 1.1 & 7.3 & &\hspace{0.28cm} (3), (5)\\
  12 & SAX J1808.4-3658 & 10 &\hspace{0.28cm} 0.12& 3.15 & &\hspace{0.28cm} (3)\\
  13 & XTE J1751-305 & 10 &\hspace{0.28cm} 1.1 & 8.0 & &\hspace{0.28cm} (3)\\
  \hline
  \label{tab1}
 \end{tabular}\\
}
\end{table*}

Using \verb HEASOFT  (version  \verb 6.24),  we first extract 16 s binned light curves in three energy bands: $4.00-18.50$, $6.00-7.50$ and $7.50-18.50$ keV from \textit{RXTE}/PCA standard 2 data for each observation. We define an ObsID \textit{hard} for which hardness ratio $H=I_{7.50-18.50}/I_{6.00-7.50}$ ($I_{7.50-18.50}$ and $I_{6.00-7.50}$ are the background uncorrected intensities in the energy bands $7.50-18.50$ keV and $6.00-7.50$ keV respectively) is  $\geq 2$ for at least $60\%$ time bins and mean $H\geq2$ \citep{Burke1}. We also choose only those observations in our work for which the background uncorrected intensity $I_{4.00-18.50}>40$ counts/s for $90\%$ time bins for achieving good statistics in the spectral analyses. We exclusively use data from PCU2, as it was always turned on. For finding the good time intervals (GTI), we exclude those time periods for which the elevation angle is less than $10^{\circ}$, offset is greater than $0.02^{\circ}$, the time since the peak of the last South Atlantic Anomaly passage is less than 30 minutes, electron contamination is greater than $0.1$ and intervals of 600s after PCA breakdown events. We also identify and exclude data contaminated by flares, dips, or thermonuclear bursts while producing the GTI files. For obtaining the light curves in three energy bands, we convert these energy bands to absolute channel number bands \footnote{\label{foot:itas2}\url{https://heasarc.gsfc.nasa.gov/docs/xte/e-c_table.html}}, which depend upon the mission epoch.

We extract background corrected spectra (using the bright background model) for those observations, which pass the criteria mentioned above, from the right and left anode chains of top xenon layer of PCU2 using the same GTI files obtained previously. Using the ftools \verb grppha, we add 0.5 percent systematic uncertainty to each bin, and the spectra are re-binned such that each bin contains a minimum of 40 counts. 

We obtain HEXTE source and background spectra using the ftools \verb hxtlcurv  and create corresponding response files using \verb hxtrsp. We use data from both the clusters A and B up to December 2005, after which the cluster A has been staring in an on-source position. Thus, we consider only cluster B for the observations done afterward.

In order to achieve a reliable and quick model fitting in \verb XSPEC  (version  \verb 12.10.1), we consider only those observations for which the background corrected PCU2 counts $>35000$, HEXTE A counts $>8000$, and  HEXTE B counts $>8000$. 

As our goal is to study the properties of Comptonised emission in X-ray binaries, we fit the spectra with a Comptonisation model including a reflection continuum and an additional Gaussian component to represent the iron K-$\alpha$ line emission. Our spectral model also includes low energy absorption by the interstellar medium. To this end, we use the \verb XSPEC  models \verb compPS  \citep{Poutanen}, which also includes the Compton up-scattered photons reflected by the accretion disc, and \texttt{ gauss}. 
\footnote{There is a number of  advanced reflection models currently available for spectral fitting, such as relxill family of models \citep{Dauser2014,Garcia2014} which include accurate account for ionization effects and relativistic smearing of the reflection features. However, the rather high low energy boundary of the \textit{RXTE}/PCA data  ($\approx 3-4$ keV) and its relatively coarse energy resolution do not permit  to utilise  advantages of these sophisticated models. On the other hand, the complexity of these models combined with above mentioned limitations of the data,  often makes the fitting procedure unstable. For this reason we choose to keep the reflection prescription based on the \texttt{ireflect}  model, implemented in \texttt{compPS}, which catches the main physical aspects of the reflection component.} 
The model \verb phabs  is used to describe the absorption column with its parameter $N_H$ being set to the Galactic value listed in the Table~\ref{tab1}. Thus, our total spectral model is \verb constant*(phabs*(gauss+compPS))  ~\citep{Burke1}. The multiplicative constant is used to take care of the calibration difference between the PCA and HEXTE instruments.  We fit the PCA and HEXTE spectra over the energy range $3-20$ keV and $20-200$ keV respectively.

While fitting the spectra with the above model, we keep the electron temperature $kT_{\rm e}$, seed photon temperature $kT_{\rm bb}$ as emitted from a multicolour disc, Compton-$y$ parameter, reflection strength $R(=\Omega/2\pi)$ ($\Omega$ is the solid angle subtended by the disc as observed from the corona) and the normalisation as free parameters in \verb compPS. We assume Maxwellian electron distribution in the Comptonising medium with a spherical geometry and a binary inclination of $45^{\circ}$ \citep{Burke1}. We do not consider those observations for spectral analysis for which $\chi^2/\nu>1.5$. Visual inspection of these spectra showed that majority of them do not belong to the classical hard state. The list of sources studied in this work, along with the number of observations considered for each of them after applying all the screening criteria mentioned above, their hydrogen column densities, distances and masses are given in the Table~\ref{tab1}.

Several of our sources (e.g., XTE J$1748-288$) lie in the close vicinity of the Galactic centre. As \textit{RXTE} has a low spatial resolution of $1^{\circ}$, spectra of some of these sources can suffer appreciable contamination due to the Galactic ridge (GR) emission. Such an emission may result in softening the spectra along with a significant contribution to the iron K$\alpha$ line. We model $3-20$ keV spectrum of the GR emission by a power law with index $\Gamma=2.15$ and a zero width Gaussian line with peak energy of 6.6 keV such that an equivalent width of $0.8$ keV is obtained \citep{Rev2003}. We also fix the flux value in this energy band using the 3.5 $\mu$m near infrared (NIR)$-$GR X-ray flux relation, as mentioned in \citet{Rev2006}. Following \citet{Burke1},  we make use of \textit{COBE}/DIRBE zodiacal subtracted mission average map for measuring the 3.5 $\mu$m flux near the source location \citep{Bennet} and also subsequently make the correction for extinction using the map of interstellar HI gas \citep{Dickey} and the extinction law by \cite{Rieke}. The contribution of GR emission is found to be appreciable (flux $>10^{-11}\ \rm erg\ \rm cm^{-2}\ \rm s^{-1}$) for the black holes: XTE J$1752-223$, XTE J$1748-288$, SAX $1711.6-3808$ and H $1743-322$, and the neutron stars: 4U $1728-34$, 4U $1724-30$, XTE J$1751-305$ and Ara X$-1$.

\begin{figure}
\begin{center}
\includegraphics[width=0.45\textwidth]{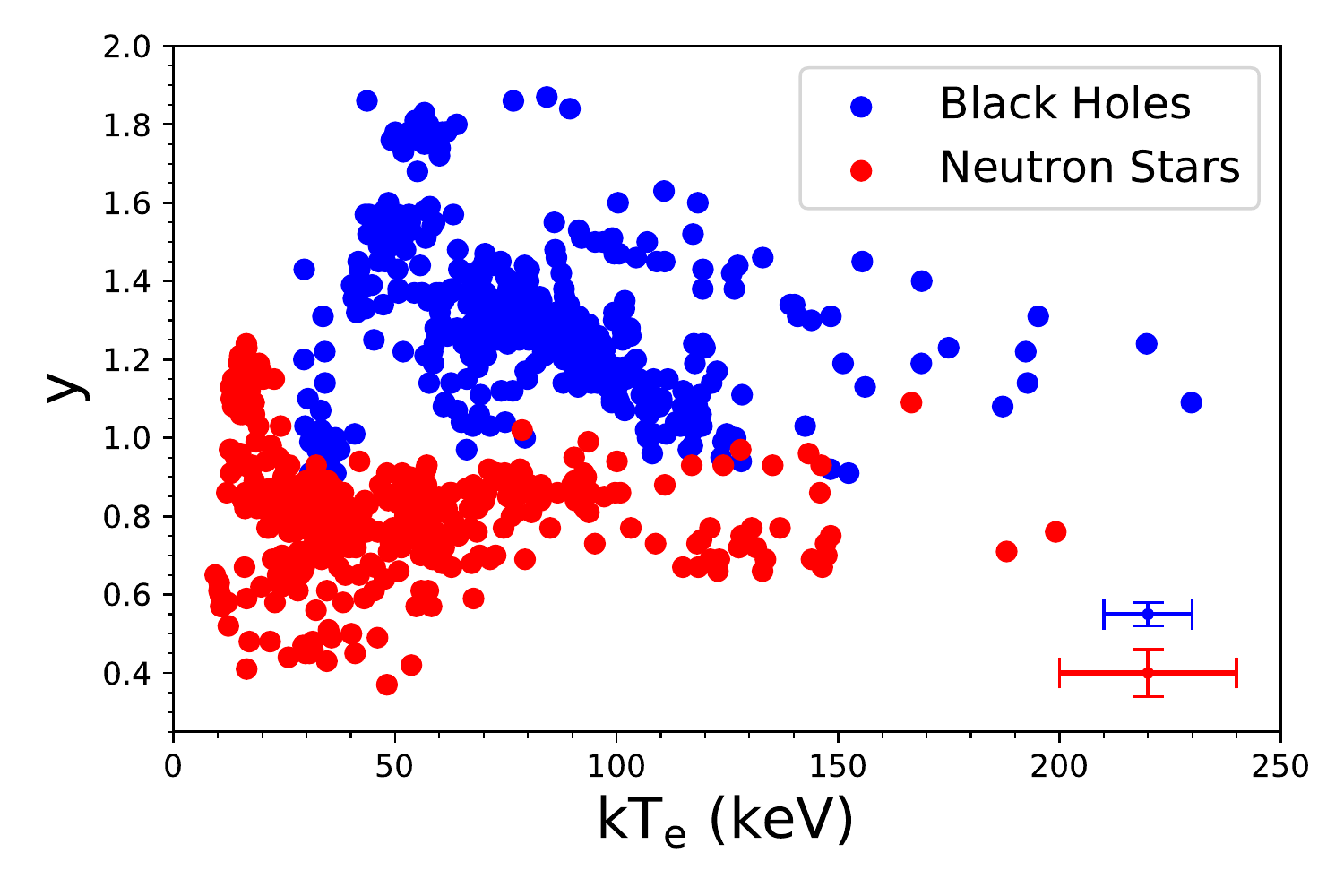}
\end{center}
\begin{center}
\includegraphics[width=0.45\textwidth]{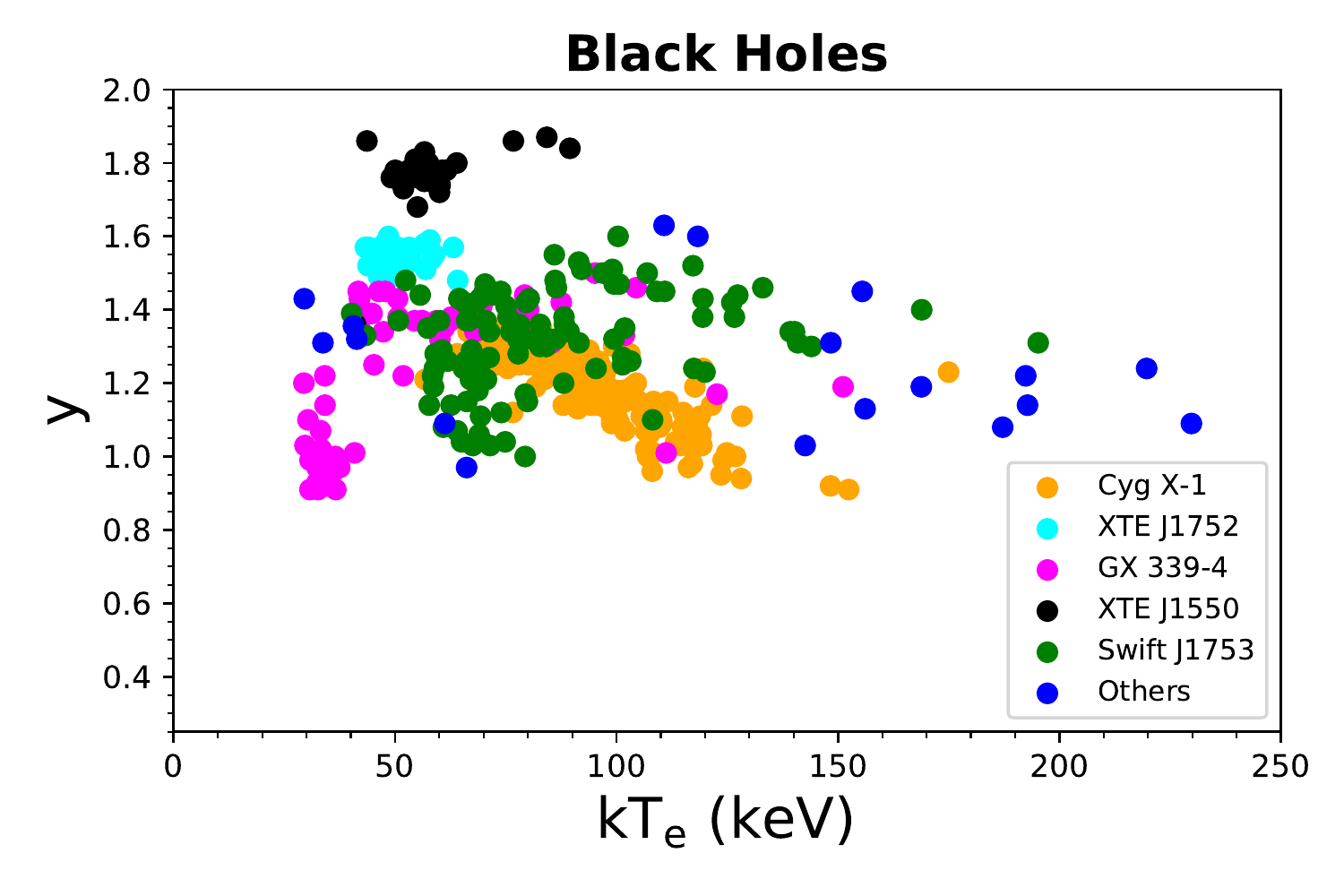}
\includegraphics[width=0.45\textwidth]{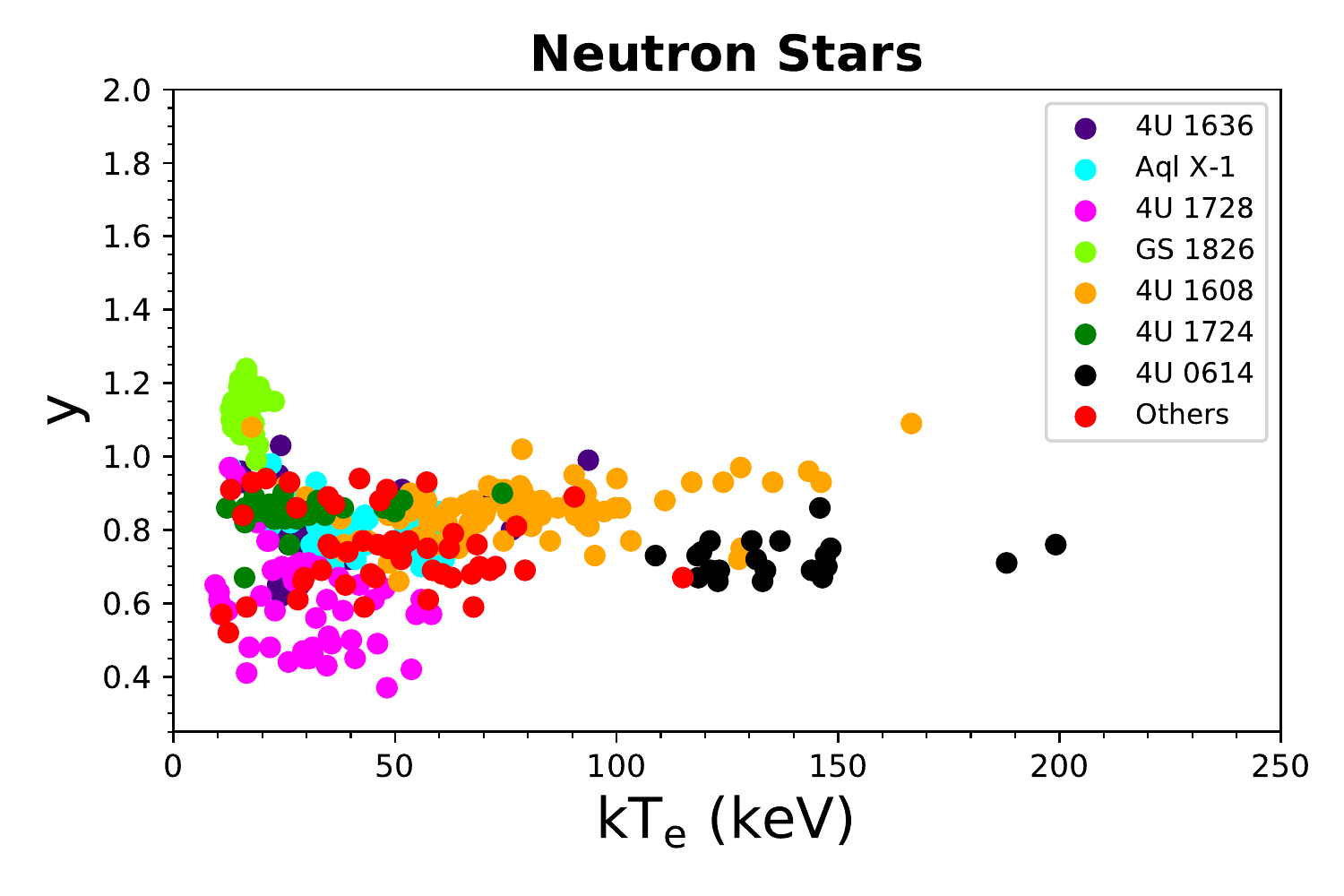}
\end{center}
\caption{The distribution of electron temperature $kT_{\rm e}$ values and the $y$ parameter values for black holes and neutron stars. In the top panel, the distribution of these parameters for all the sources considered in this study are depicted. The distribution of data on the $y$ -- $kT_{\rm e}$ plane for multiple observations of different sources -- black holes (middle panel) and neutron stars (bottom panel) are also depicted here. The sources are colour coded, as explained  in the legend.  Typical error bars for black hole and neutron star systems are shown in the top panel. As one can see, different sources tend to occupy different sub-regions in the diagram; however taken collectively, black holes and neutron stars occupy well separated distinct regions on this plane (top panel). See section \ref{res1} for more details. \label{fig1}}
\end{figure}

\section{Results and Discussion}\label{result}
\subsection{Correlations between parameters of Comptonised spectra}\label{res1}
The best fit values of the Compton $y$-parameter and electron temperature $kT_{\rm e}$ are shown in Fig.\ref{fig1}. We find that the black holes (BHs) and neutron stars (NSs) occupy distinctly different regions in the $y-kT_{\rm e}$ plane, with  the BHs tending to have larger $y$-parameter and higher electron temperature $kT_{\rm e}$.

As we see in Fig. \ref{fig1},  the boundary  between BH and NS systems has a complex shape, which is why  their marginalised one-dimensional distributions over $y$-parameter and $kT_{\rm e}$ overlap, especially for the latter (see Fig. \ref{fig5} and discussion in section \ref{res3}). However, when considered on the two-dimensional $y-kT_{\rm e}$ plane, there is virtually no overlap between the two types of compact object. For example, the electron temperature of the Comptonising media in NSs 4U $1608-52$ and 4U $0614+09$ reach values greater than $100$ keV, unlike the other NSs in our sample. However, their corresponding $y$-parameter values are in the $y\sim 0.7$ range, i.e., significantly less than that of BHs with a similar electron temperature. Similarly, although the $y$-parameter of the spectra for the NS GS $1826-238$ can reach values of $y\sim 1.2$, their corresponding electron temperatures are less than $20\ \rm keV$ (see Fig. \ref{fig1}). We note that the anomalous hardness of 4U $1608-52$ was noticed  in \citet{Burke1}; in our analysis, due to a larger sample, it reveals itself yet more graphically. For the same reason, while \citet{Burke2} found 4U $0614+09$ as a quite normal NS system, in our sample, it shows the highest $kT_{\rm e}$ among NSs. (We note, however, that  the two rightmost points for this source in the bottom panel in Fig. \ref{fig1} have statistical errors of $\sim 40$ keV.) This shows that the size of the sample is an important factor in a comprehensive sampling of the parameter space of Comptonisation  in X-ray binaries. 

\begin{figure}
\begin{center}
\includegraphics[width=0.45\textwidth]{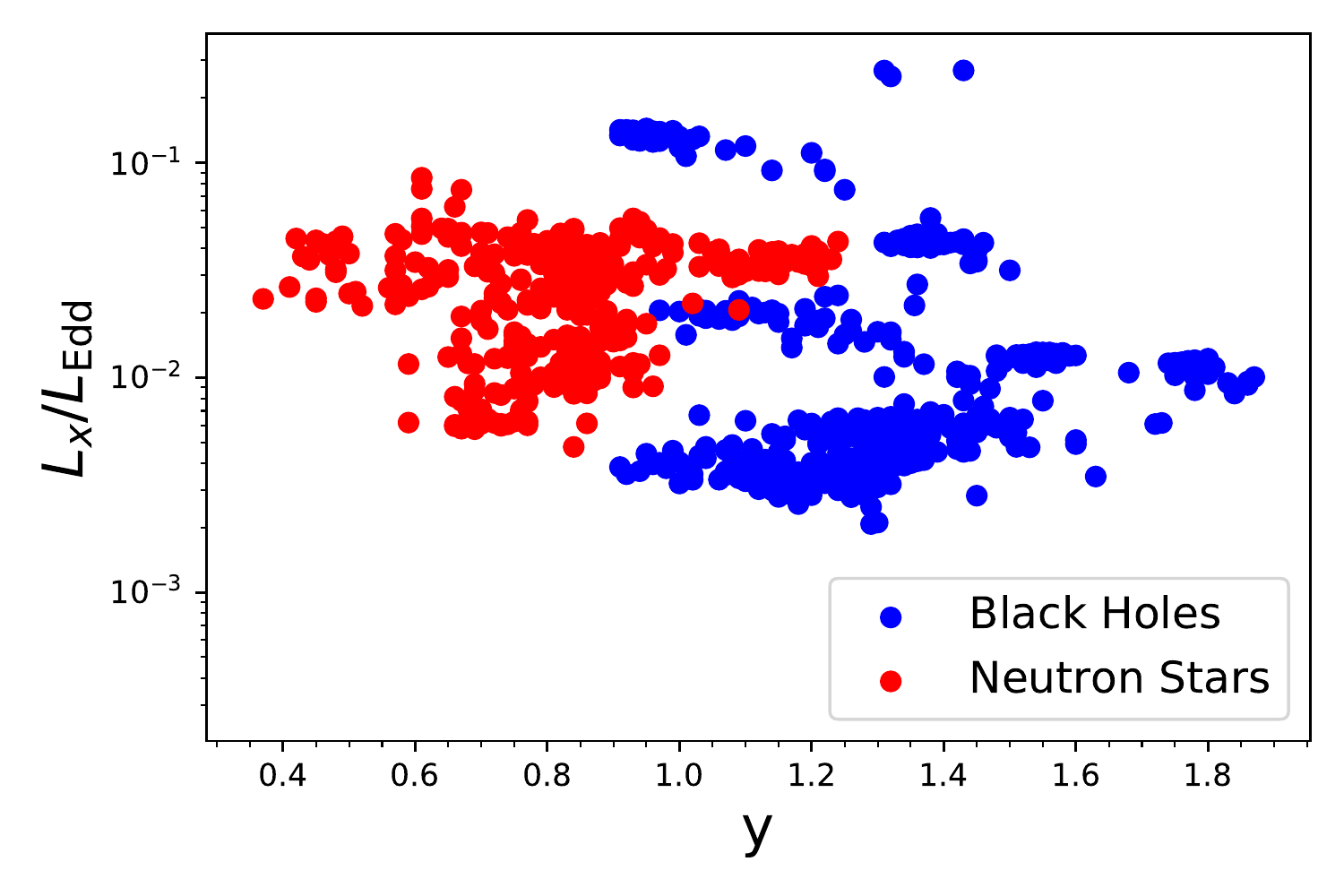}
\end{center}
\begin{center}
\includegraphics[width=0.45\textwidth]{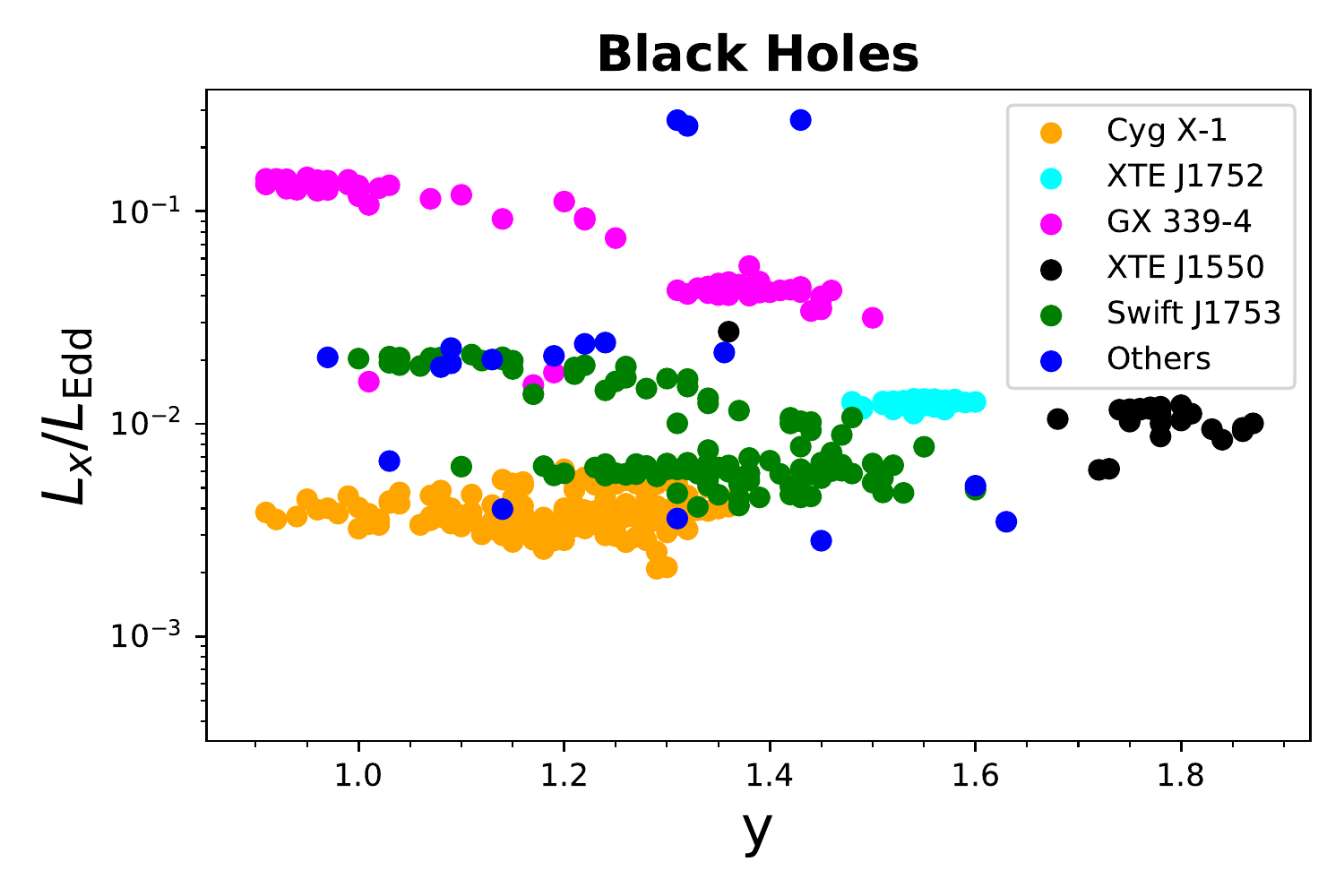}
\includegraphics[width=0.45\textwidth]{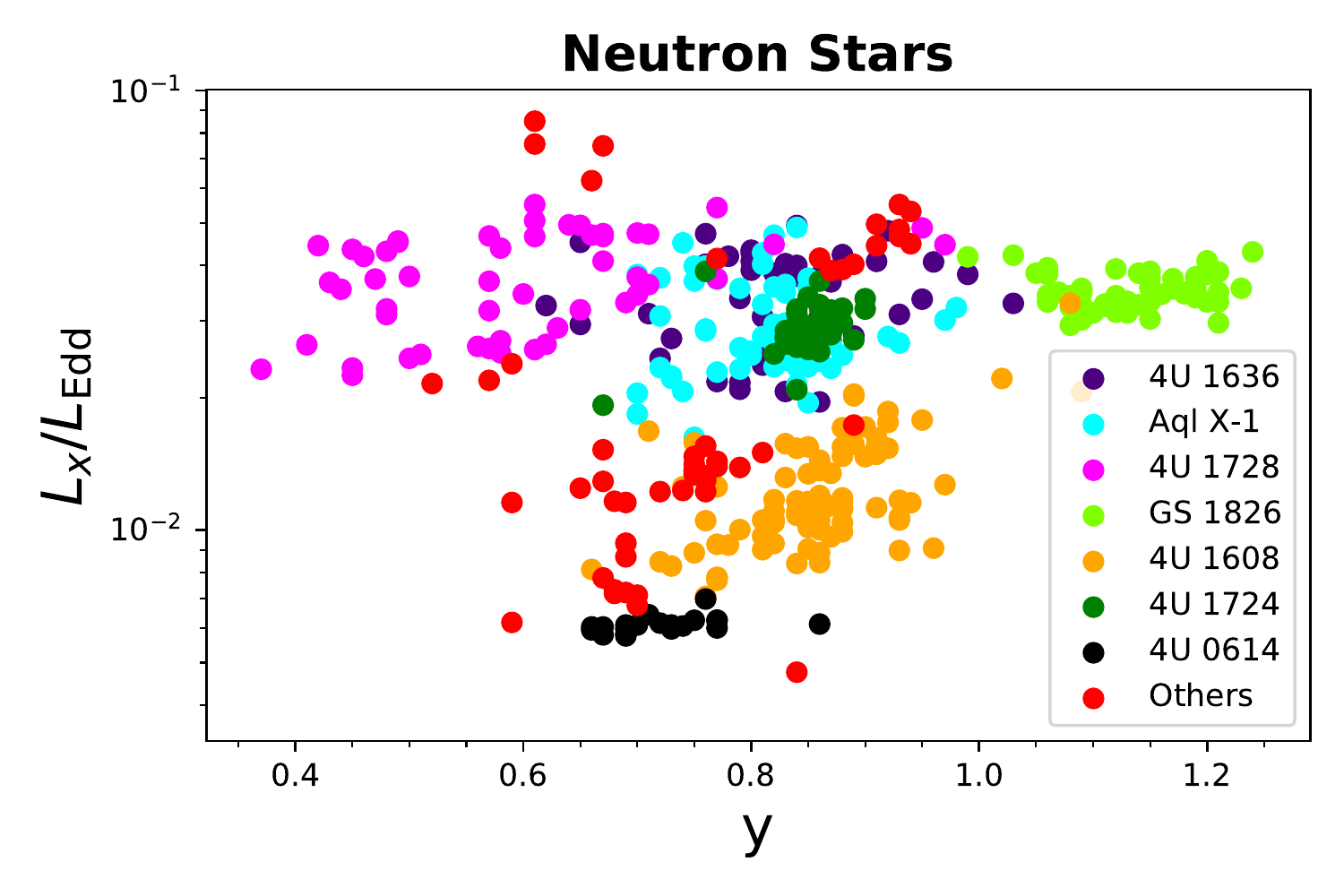}
\end{center}
\caption{The distribution of 3--200 keV luminosity and the $y$ parameter values for black holes and neutron stars. In the top panel, the distribution of these parameters for all the sources considered in this study are depicted. The two lower panels show respective distributions for black holes and neutron stars with the data for different sources colour coded, as explained  in the legend. Note that the three panels have different axes scales. See section \ref{res1} for more details. \label{fig_lx}}
\end{figure}

Interestingly, different spectra of a given source tend to group together rather than uniformly fill the entire BH or NS domain on the $y-kT_{\rm e}$ plane (Fig. \ref{fig1}, two lower panels). Partly, this may be possibly explained by the fact that we fixed the binary inclination on the same fiducial value of $45\degr$, while due to asymmetries in the disk-corona configuration, there is some mild viewing angle dependence in the Comptonised spectra. However, this is unlikely to explain the full range of values in Fig. \ref{fig1}. The reason (or parameter responsible) for such behaviour is unclear and needs further  investigation \citep[cf.][]{Gladstone, heil2015}. We note, however, that this parameter is not the mass accretion rate (as traced by X-ray luminosity) as it is confirmed with the plots in Fig.\ref{fig_lx}. 

The spectral model includes reflected component accounting for the result of reprocessing of the Comptonised spectrum by the accretion disc and other optically thick media (if any) located in the vicinity of the Comptonising region. The amount of reprocessed emission in the total spectrum is characterised by the reflection strength $R=\Omega/2\pi$, where $\Omega$ is the solid angle subtended by the disc as observed from the corona. With the exception of a small number of spectra, the best fit values of $R$ are all smaller than unity, as it should be expected in commonly considered geometries of the accretion flow, where the main reprocessing site is the accretion disc.

It was earlier reported \citep{Gladstone} that  atoll sources and millisecond pulsars exhibit some differences in their behaviour on the colour-colour diagram during the hard/soft transition,  on basis of which they can be divided into two groups: the sources (e.g., 4U $1636-536$, 4U $1728-34$) which make a diagonal track on the colour-colour diagram and the sources (e.g., 4U $1608-52$, Aql X-1) which make a vertical track. Our sample includes NSs from each subclass, however, we do not find any difference in their Comptonisation parameters. 

\subsection{Compton amplification factor}
\label{res2}

The Compton amplification factor is defined as a ratio of the luminosity of the Comptonised component to the luminosity  of the soft seed photons. As a proxy to the former, we used \verb compPS  model luminosity in the $3-200$ keV band, corresponding to the energy range where X-ray data was fit. 
We calculate the seed photon flux as flux of the \verb compPS  model, for each best-fit model setting $\rm cosIncl=-0.5$ and $\tau=10^{-4}$. With these parameters, \verb compPS  model spectrum equals to the input spectrum of seed photons. Its flux is then computed with \verb XSPEC  flux command in an energy range from 1 eV to 10 keV.

We note that due to the limited energy range used for spectral fitting, the so defined $A$ is some approximation to the true value of the Compton amplification factor.

\begin{figure}
\begin{center}
\includegraphics[width=0.42\textwidth]{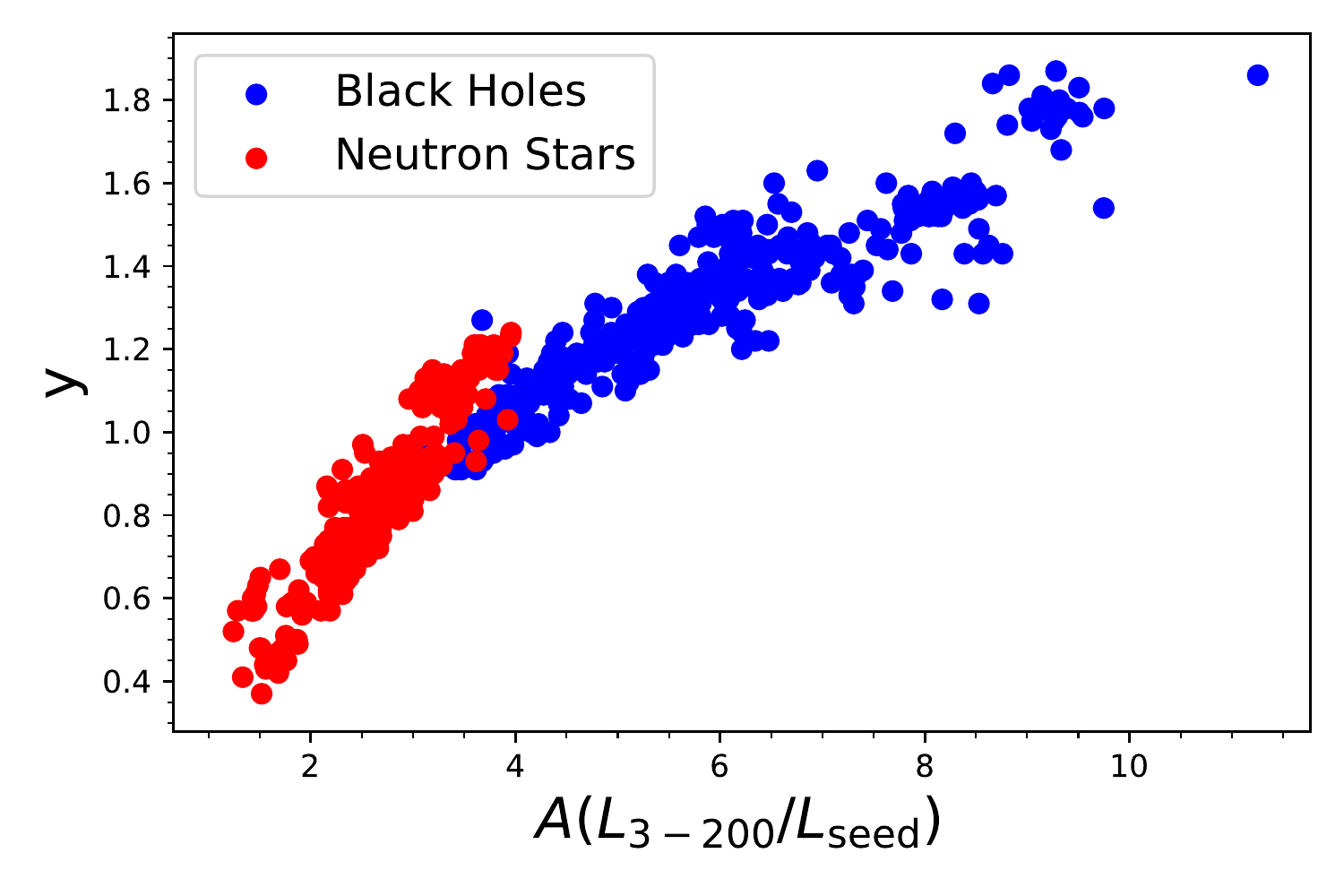}
\includegraphics[width=0.42\textwidth]{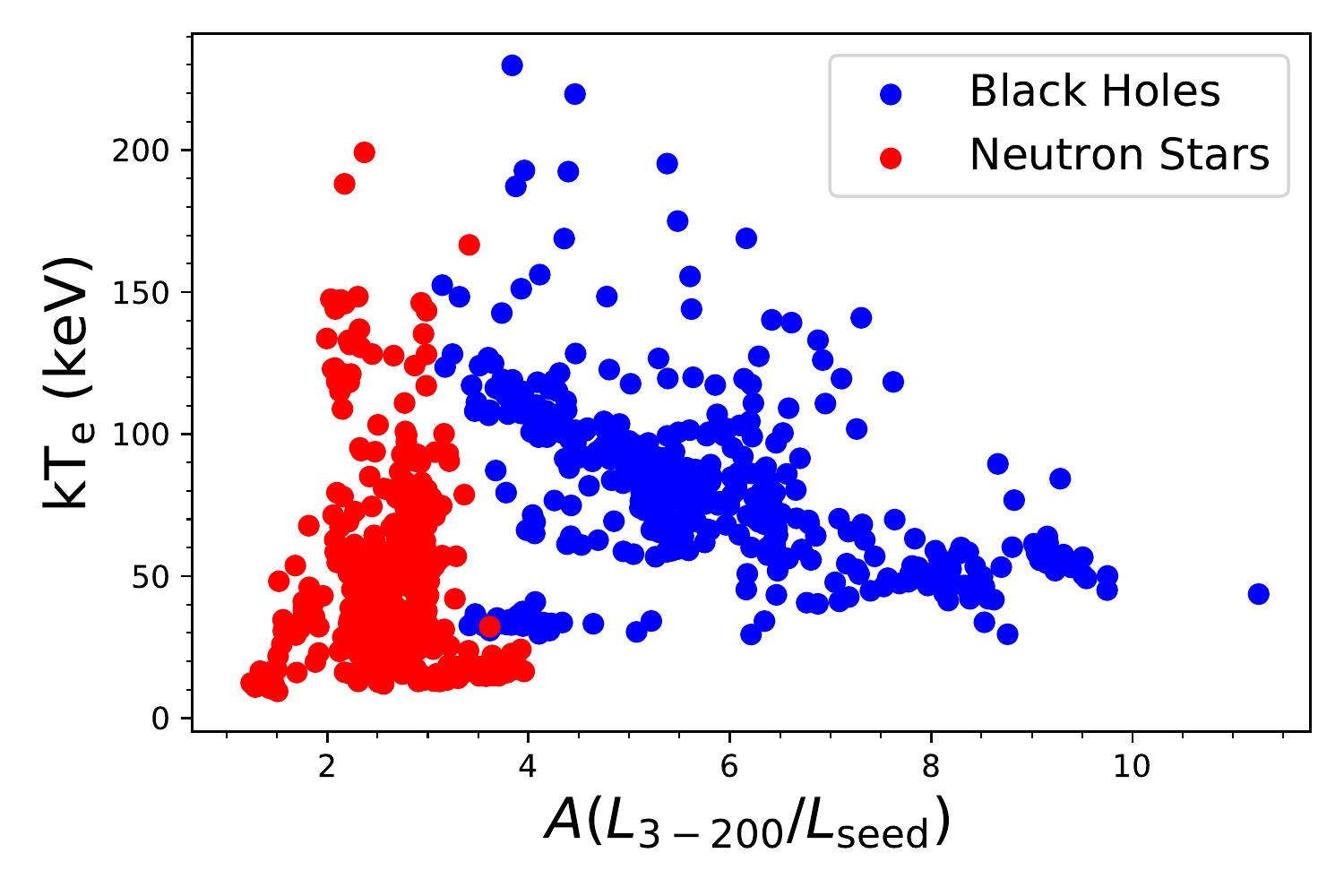}
\caption{Relation between the Compton amplification factor and $y$-parameter (top panel) and $kT_{\rm e}$ (bottom panel) for black holes and neutron stars.  See section \ref{res2} for more details.\label{fig4}.}
\end{center}
\end{figure}

The Compton amplification factor $A$  is not an independent parameter of the spectral fit and is derived from the best fit parameters. It characterizes the energy balance in the Comptonising region and is closely related to the parameters of the Comptonised spectrum, with larger $A$ corresponding to larger $y$ and flatter spectra \citep[e.g.][]{gilfanov2000}. As discussed in the following section, one should  expect that in BH systems, $A$ is larger than in NS systems. The distribution of BH and NS systems in the $A-y$ and $A-kT_{\rm e}$ planes is shown in Fig.\ref{fig4} confirming this expectation. 

It is interesting to note that the distribution of points on the $A-y$ plane is essentially one-dimensional. In the low-left corner of the plot (low $A$ and $y$), this is a consequence of some degeneracy between these parameters in the low $A$ regime (see, for example, section 2.8 in \citealt{gilfanov2010}). In the upper-right end of the dependence, however, the range of permitted values of $A$ for the given $y$ is broader than it is observed. Therefore, the narrowness of the distribution may, in principle,  reflect some real non-trivial correlations between Comptonisation parameters in BHs. Further investigation of this behaviour is hampered by the limited low energy coverage of the data used in this work.

\subsection{Comptonisation in BH and NS systems -- evidence for the BH event horizon and other implications}
\label{res3}

As one can see from Fig.~\ref{fig1} and \ref{fig4}, there is a clear dichotomy between BHs and NSs in the Comptonisation parameter space with virtually no overlap between the two types of compact object. BHs have larger values of $y$-parameter and their electron temperature distribution is shifted towards higher $kT_{\rm e}$ values. These results confirm and expand,  with a much larger sample of sources and higher statistics,  earlier findings of \citet{Burke1, Burke2}. An observational manifestation of the  dichotomy in Comptonisation parameters of BHs and NSs is the fact that the hard state spectra of the NSs are generally softer than  spectra of BHs, as it was initially pointed out in \citet{Syunyaev91} based on the experience of  MIR-KVANT and GRANAT observations of Galactic compact sources. 

The difference in Comptonisation parameters in BH and NS systems  can be understood as a result  of more efficient cooling of electrons near a NS due to the presence of the more abundant cooling agent -- soft  photons produced by the NS surface.  The NS emission is powered by the remaining kinetic energy of the accreting material decelerated upon reaching the slowly rotating NS, and may surpass  other sources of seed photons for Comptonisation, such as the accretion disc. Such an additional source of soft radiation is not available for BHs, as in their case, the remaining kinetic energy of the accreting material is advected inside the event horizon, without being converted to radiation. The additional cooling for the Comptonising region present in the case of accretion onto an NS leads to a decrease of the Compton $y$-parameter,  electron temperature, and Compton amplification factor, cf. Fig.\ref{fig1},\ref{fig4}. 

\begin{figure}
\begin{center}
\includegraphics[width=0.42\textwidth]{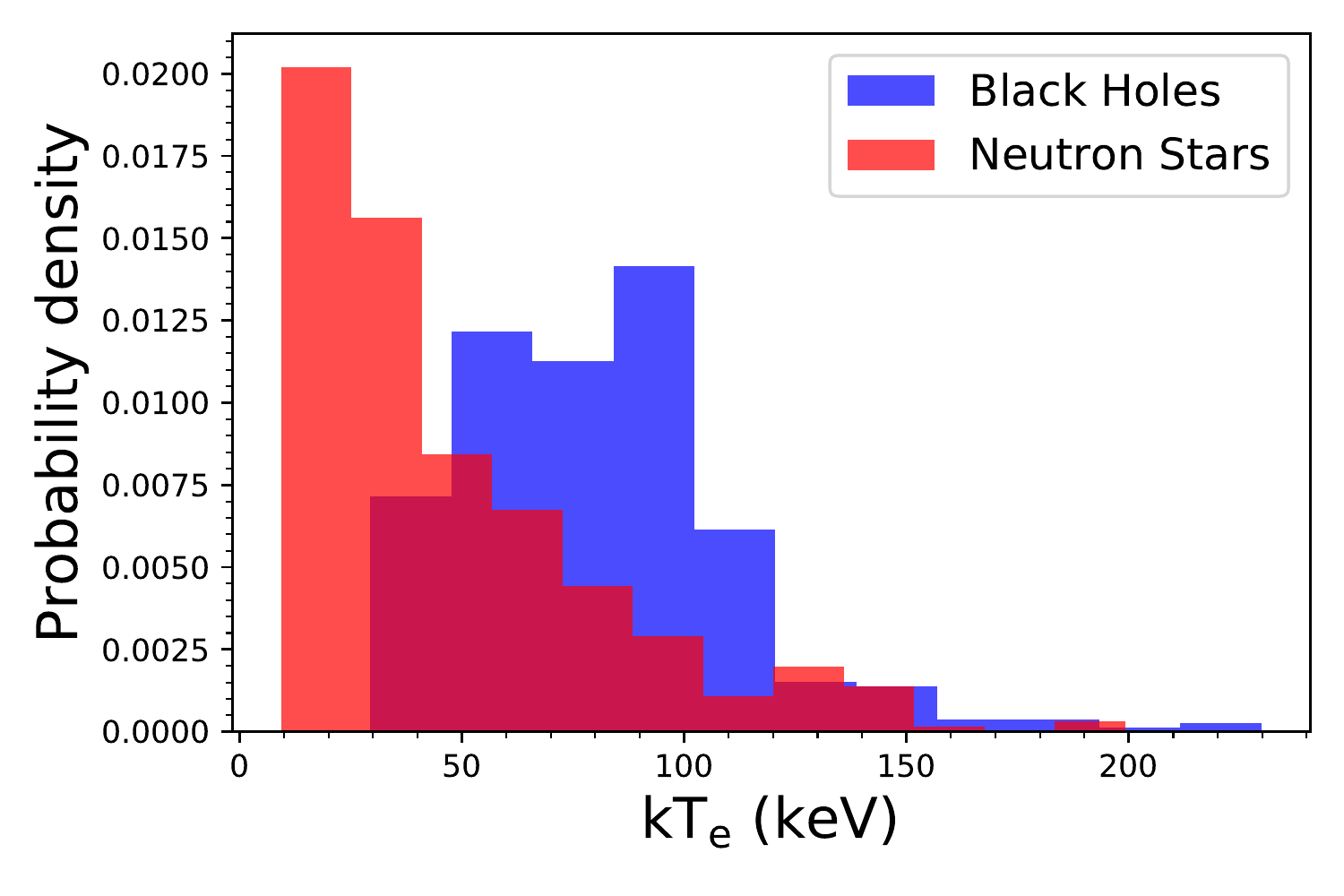}
\end{center}
\begin{center}
\includegraphics[width=0.42\textwidth]{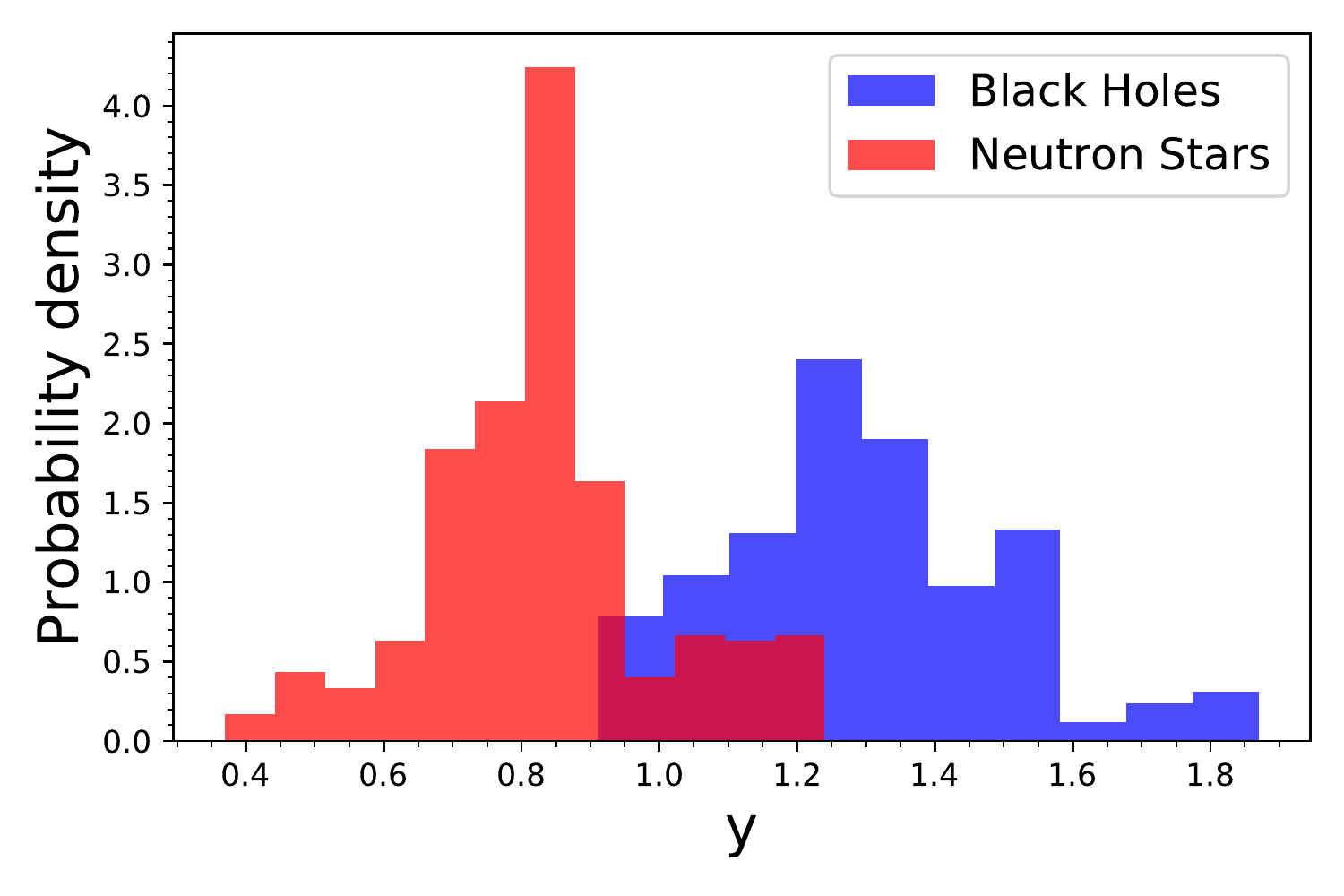}
\includegraphics[width=0.42\textwidth]{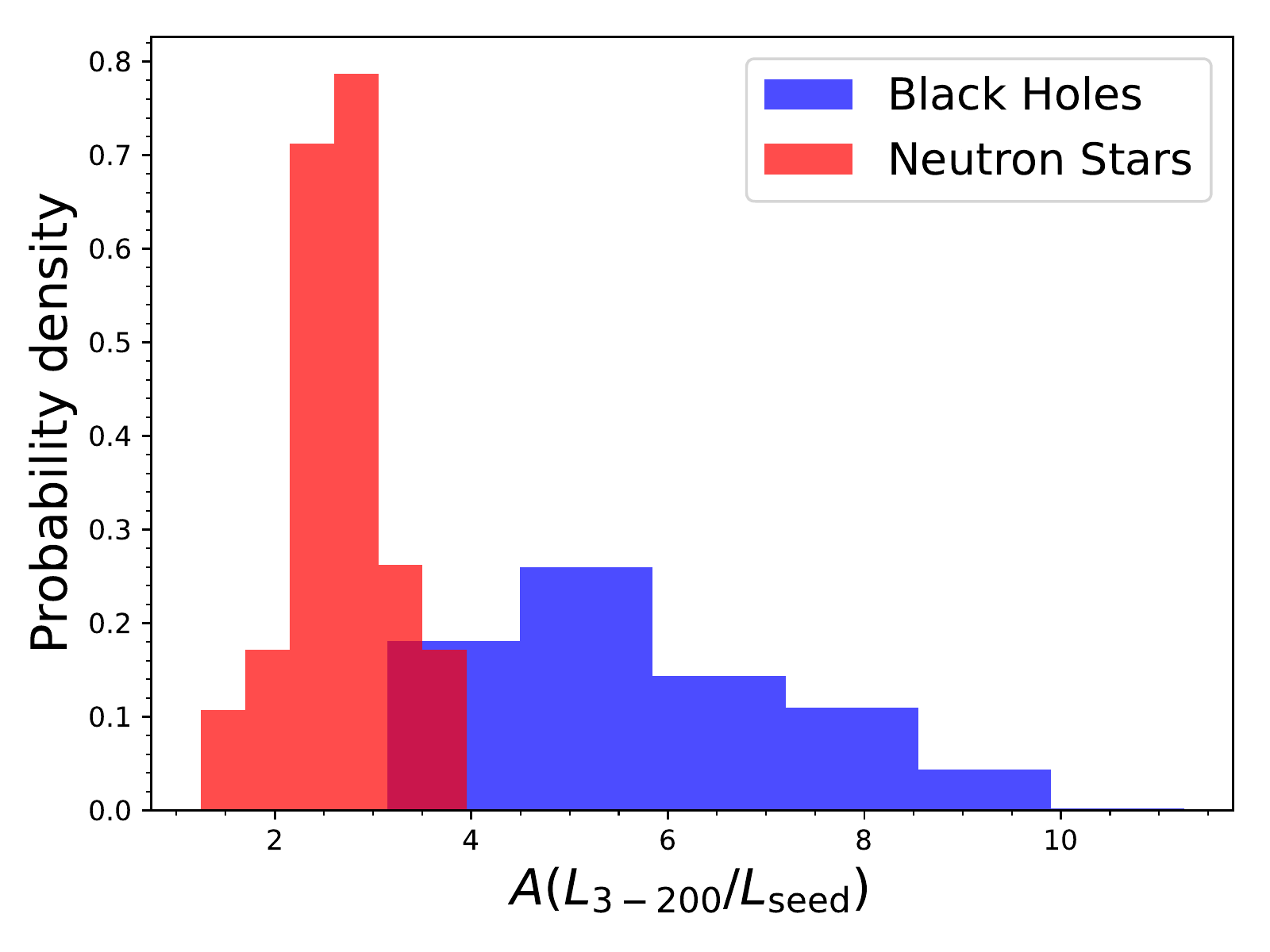}
\caption{Marginalised probability density distributions of Comptonisation parameters. The marginalised probability density distributions of electron temperature $kT_{\rm e}$ (top panel), Compton $y$ parameter (middle) and Compton amplification factor $A$ (bottom) for black holes and neutron stars are depicted here. See section \ref{res3} for more details.\label{fig5}}
\end{center}
\end{figure}

The marginalised one-dimensional  probability distributions of $kT_{\rm e}$, $y$-parameter, and $A$ are shown in Fig. \ref{fig5}. They all have apparently different shapes for BHs and NSs. To quantify this, we performed the Kolmogorov-Smirnov (KS) tests on BH and NS distributions for each parameter. We obtained the KS-statistic value (p-value) of $0.80$ ($p\approx1.1\times10^{-121}$), of $0.5$  ($p\approx8.1\times10^{-47}$) and $0.92$ ($p\approx2.2\times10^{-160}$) for  $y$-parameter, $kT_{\rm e}$ and  the amplification factor $A$. These results confirm the visual impression from Fig. \ref{fig5}  that the BH and NS distributions  differ with an extremely high statistical significance (e.g. for $y$ and $A$ it is  equivalent to $> 20\sigma$ for a Gaussian distribution).

Because of the curved shape of the boundary between BHs and NSs in the $y-kT_{\rm e}$ plane, there is a considerable overlap between their  $y$ and, especially $kT_{\rm e}$ distributions. Nevertheless,  their mean values and quantiles are quite different for BHs and NSs. The mean of $kT_{\rm e}$ for our NS sample is $\approx47$ keV, and its 80\% quantile is $16-93$ keV, while BHs are characterised by the mean of $\approx80$ keV and $80\%$ quantile of $45-117$ keV. For Compton $y$-parameter,  mean values are  $\approx 0.83$ and $\approx 1.3$ for NSs and BHs, their 80\% quantiles :  $0.64-1.09$ and  $1.03-1.55$ respectively.

If one needed to devise a one-parameter test however, the Compton amplification factor $A$ would be the primary  parameter discriminating between BHs and NSs (Fig. \ref{fig4},\ref{fig5}). Indeed, the $A$ distributions have mean values of $\approx 2.7$ and $\approx 5.8$ and 80\% quantiles of  $2.08-3.32$ and  $3.94-8.17$ for NSs and BHs respectively. This should have been expected, as  by definition, $A= L_{\rm tot}/L_{\rm seed}$ where $L_{\rm tot}$ is the total X-ray luminosity of the source and $L_{\rm seed}$ is the luminosity of seed photons for Comptonisation.  Following the notation of \citet{Burke1} (their section 5.2), one can write for an NS: $L_{\rm seed}=f_1\,L_{\rm disc}+f_2\,L_{\rm NS}$. Thus, at the same total luminosity $L_{\rm tot}$, in an NS X-ray binary $L_{\rm seed}$  should be  larger due to the additional soft radiation  $L_{\rm NS}$ from the surface of the NS, which drops the value of $A$. For BHs, having no hard surface, seed photons, in the simplest scenario, are generated by the accretion disc only, $L_{\rm seed}=f_1\,L_{\rm disc}$, and $A$ is correspondingly larger. 

One can use these simple considerations to roughly estimate the fractions of the kinetic energy of accreting material released in the Comptonising corona and dumped onto the surface of the NS. Using the mean value of $A\approx 2.7$ determined  for NSs in our sample and eq.(7) from \citet{Burke1} we obtain $W_{\rm corona}/W_{\rm NS}\approx 1.7$. Therefore about $\sim 2/3$ of the energy of accreting material is released in the Comptonising corona around the NS, with the remaining $\sim 1/3$ being released on the NS surface. These numbers are consistent with those derived in \citet{Burke1}, who obtained the corona fraction of $\sim 1/2-2/3$. 
The general conclusions from both calculations are that (i) the corona is radiatively highly efficient and (ii) the NS makes significant or dominant contribution to the seed photon supply for Comptonisation.

\section{Summary and Conclusion}\label{summary}
A black hole lacks a hard surface and is confined by an invisible boundary, an event horizon, a definitive proof of which is a holy grail of modern physics and astronomy. A signature of this horizon could be identified, if the X-ray radiation from accreting stellar-mass black holes is compared with that from accreting low magnetised neutron stars, which have hard surfaces. Here, we investigate how the additional X-ray photons emanating from the surface/boundary layer of neutron stars can affect the Comptonisation process in the hard state, and how such interaction is manifested on the distribution of the properties of Comtonising components: the corona temperature $kT_{\rm e}$, the Compton $y$ parameter, and the Compton amplification factor $A$. We analyse the hard X-ray spectra from 11 accreting black holes and 13 atoll neutron stars as observed with \textit{RXTE}/PCA and \textit{RXTE}/HEXTE instruments. We find a clear dichotomy in the distributions of black holes and neutron stars on the $y-kT_{\rm e}$ plane (Fig. \ref{fig1}) and in the distributions of the Compton amplification factor (Fig. \ref{fig4}). The values of the Compton $y$-parameter and the amplification factor $A$, in general, take higher values for black holes than neutron stars. However, there is a significant overlap between the marginalised one-dimensional distributions of corona temperatures for black holes and neutron stars, the neutron stars, on average, have lower values of corona temperature than the black holes. 

Thus, our study reveals a clear separation between Comptonisation characteristics of accreting black holes and neutron stars, which  is  a consequence of the lack of the hard surface in black holes. Our findings establish a new method of determining the nature of the compact object in X-ray binaries through the broad-band X-ray spectroscopy. Previously, several techniques for identifying  stellar-mass black holes in X-ray binaries have been proposed, based on dynamical measurements  of the mass of the accretor  via binary system mass function determination \citep{Remillard} or measurement of characteristic time scales in the vicinity of the compact object through aperiodic variability of its X-ray emission \citep{mikej}, or through the shape of the correlation between the QPO frequencies and the power law  slope of X-ray spectrum \citep{tit2006}. It was also demonstrated that black hole X-ray novae typically have a 1--2 orders of magnitude lower quiescent luminosity than neutron star transients \citep{narayan1997}, and it was proposed on theoretical grounds that black hole X-ray binaries never exhibit  Type I X-ray bursts \citep{narayan2002}.  The latter two suggestions  pioneered the use of the signatures of the black hole event horizon for the compact object diagnostics. In this paper, we propose a method that can be employed for the compact object diagnostics given a single broad-band X-ray spectrum of a source in the hard state.

Although the first supermassive black hole has already been imaged with the Event Horizon Telescope using electromagnetic radiation \citep{Event}, such observations are still far from becoming a routine.  Moreover,  it may not be possible to image a stellar-mass black hole in the  foreseeable future, because of its several orders of magnitude smaller angular size. Therefore, the proposed method should be useful to detect the signatures of event horizon of  stellar-mass black holes using the observations of their electromagnetic emission. This would be particularly important to  probe the  regime of strong gravity, as stellar-mass black holes produce several orders of magnitude larger space-time curvature compared to those of supermassive black holes. Although merging stellar-mass black holes have been detected  via their gravitational wave emission \citep{Ligo}, those are very short transient events and can be studied only once during the merger. Accreting black holes in low mass binaries, on the contrary, are stable, long-lived systems, for which the results can be refined in the future with better data and more advanced techniques.

\section*{Acknowledgements}

This research has made use of data, software, and/or web tools obtained from the High Energy Astrophysics Science Archive Research Center (HEASARC), a service of the Astrophysics Science Division at NASA/GSFC.  This work was partially supported by the Russian Science Foundation, project \#19-12-00369.  S. Banerjee is thankful to Lavneet Janagal, Dattaraj Dhuri and Prasanta Kumar Nayak for downloading a significant part of the data used in this work. 
We thank the referee Prof. Chris Done for her insightful comments and suggestions.

\section*{Data Availability}

The  observational  data  used  in  this  paper  are  publicly available  at HEASARC (\url{https://heasarc.gsfc.nasa.gov/}). Any  additional  information  will  be  available upon reasonable request.




\bibliographystyle{mnras}
\nocite{*}

\bibliography{mnras.bib} 



\bsp	
\label{lastpage}
\end{document}